\documentclass[twocolumn,prstab]{revtex4-1}
\usepackage{graphicx}% Include figure files \usepackage{dcolumn}% Align table columns on decimal point \usepackage{bm}% bold math%\usepackage{epsfig,Here} \usepackage{lineno} \linenumbers

\begin{document}
\title{Halo Coupling and Cleaning by a Space Charge Resonance in High Intensity Beams}

\author{Ingo Hofmann}
 \altaffiliation[Also at ]{Helmholtz-Institut Jena,
Helmholtzweg 4, 07743 Jena, Germany}%Lines break automatically or can be forced with \\
\affiliation{Gesellschaft f\"{u}r Schwerionenforschung (GSI),
Planckstr. 1, 64291 Darmstadt, Germany}  \email{i.hofmann@gsi.de}

%\date{\today}

\begin{abstract}

 We show that the difference
resonance driven by the space charge pseudo-octupole of
high-intensity beams not only couples the  beam core emittances;
it can also lead to emittance exchange in the beam halo, which is
of relevance for beam loss in high intensity accelerators. With
reference to linear accelerators the ``main resonance''
$k_z/k_{x,y} =1$ (corresponding to the Montague resonance
$2Q_x-2Q_y=0$ in circular accelerators) may lead to such a
coupling and transfer of halo between planes. Coupling of
transverse halo into the longitudinal plane - or vice versa - can
occur even if the core (rms) emittances are exactly or nearly
equal. This halo argument justifies additional caution in linac
design including consideration of avoiding an equipartitioned
design. At the same time, however, this mechanism may also qualify
as active dynamical halo cleaning scheme by coupling a halo from
the longitudinal plane into the transverse plane, where local
scraping is accessible. We present semi-analytical emittance
coupling rates and show that previously developed linac stability
charts for  the core   can be extended - using the longitudinal to
transverse halo emittance ratio - to indicate additional regions
where halo coupling could be of importance.

\end{abstract}

\maketitle \vspace{1cm}

%\newpage            \section{Basic Features} \label{sec:2}
\setcounter{page}{1}
\section{INTRODUCTION} \label{sec:1}
 The possibility of manipulating the 6D phase space
distribution of particle beams by exchanging emittances between
planes can be important for optimum performance as well as for
minimizing beam loss. In the context of free electron lasers, for
example, emittance exchange between transverse and longitudinal
planes has been proposed to take advantage of very small   source
longitudinal emittances. For this purpose  dispersion-based
concepts for multi-GeV electron beams involving chicanes,
transverse RF cavities as well as quadrupoles have been
suggested~\cite{xiang-chao2011}. For (partially stripped) ion
beams the proposal was made recently that a beam with equal
transverse emittances can be effectively transformed into a more
flat beam matched better to subsequent ring injection, where this
rotator includes normal quadrupoles, a stripper, chicane and skew
quadrupoles~\cite{groening2011}.

For circular machines the idea of employing lattice nonlinear
resonances  to work on beam tails and halo cleaning was proposed
some time ago~\cite{chao1976}. With reference to the CERN
Intersecting Storage Ring the suggestion   was made to use a
5$^{th}$ order resonance to ``pump'' protons into the tail up to
the aperture or a distant scraper. This would allow to clean the
halo dynamically rather than by applying - as usual -  a scraper
close to the beam core. In a different context coupling between
tails was employed experimentally at the CERN Proton Synchrotron,
where insufficient (transverse) Landau damping of head-tail modes
in one plane could be successfully enhanced by linear coupling
with the tail distribution in the other plane
plane~\cite{metral1998}.

In high intensity accelerators any active concept addressing
emittance exchange needs to deal with space charge and possible
side effects of it. In our study we consider the intrinsic space
charge pseudo-octupole - always present in a non-uniform spatial
distribution - as driving term of a nonlinear difference resonance
and explore its possible role in an active emittance coupling
scheme. Core and halo emittance coupling is an important issue for
high intensity linacs, where trade-offs between longitudinal
acceptance and synchronous phase require careful consideration
(see, for example, Ref.~\cite{fu}).

 In linear accelerators this space charge difference
resonance is known as ``main resonance'' with the equivalent
resonance condition in linac notation $k_z/k_{x,y} =1$. For a
discussion of its theory see
Refs.~\cite{epac2002,hofmann2006,hofmann2012}; furthermore
Ref.~\cite{groening2009b} for the first experimental observation
of the linac main resonance at the GSI UNILAC. For reference we
mention that in circular accelerators this mechanism is better
known as ``Montague resonance''~\cite{montague} with the
corresponding difference resonance condition $2Q_x - 2Q_y=0$. For
a detailed experimental study of the Montague resonance in the
CERN Proton Synchrotron see Ref.~\cite{metral-epac}.

  In discussions of high intensity linac beams it is often assumed that
  resonant interaction with a space charge driven
  nonlinearity primarily affects   the beam core.
  For the latter it is
 assumed that rms emittance transfer between longitudinal and transverse planes should be
avoided.
 This can be achieved either by using a
priori ``equipartitioned'' beams, or simply by avoiding the
corresponding resonance condition and work with
non-equipartitioned beams and a higher flexibility of lattice
tuning. The latter option has found new interest for H$^-$-linacs,
where avoiding the risk of intrabeam stripping losses~\cite{ibs}
is helped by reduced transverse focusing and removing the
constraint of an equipartition requirement.

 The main point of this
study is to explore whether the   emittance coupling concepts
describing primarily the beam core can be extended to describe
coupling effects in the  beam halo - whatever the origin of it may
be. Along this line  we also examine the possibility  of employing
space charge and the core-halo coupling in a strategy for active
  transfer of longitudinal halo into the transverse plane, where it can be scraped
   in a well-controlled way.

Our study is carried out with the TRACEWIN code in an idealized
transport lattice with RF focusing, hence ignoring acceleration
and all additional complexity of a specific linac lattice. In
Section~\ref{sec:2} we present some details of the ideal model; in
Section~\ref{sec:3} results for dynamical crossing of the main
resonance with/without halo; in Section~\ref{sec:4} we present
results for fixed tunes; in Section~\ref{sec:5} analytical
scalings are summarized; in  Section~\ref{sec:6} we introduce
extended stability charts, with concluding remarks in
Section~\ref{sec:7}.

\section{Lattice, beam core and halo definitions}\label{sec:2}
\begin{figure}[h]
\centering\resizebox{0.45\textwidth}{!}{\includegraphics{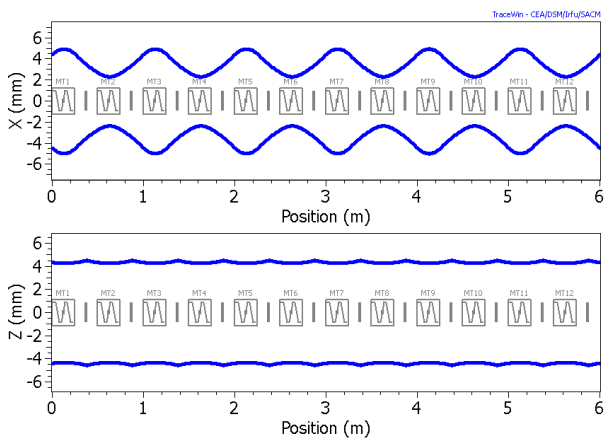}}
\caption{TRACEWIN envelopes for reference transport channel with
RF focusing gaps.} \label{transport}
\end{figure}

Without lack of generality we employ an ``idealized accelerator''
lattice with a periodic FODO channel (85 cells, with cell length
arbitrarily chosen as 1 m) with linearly varying transverse phase
advance and - for simplicity - a single RF kick per cell to
generate longitudinal focusing with a fixed phase advance of
74$^0$ per cell throughout the paper (Fig.~\ref{transport} only
showing first six cells).
The current is chosen such as to yield a transverse tune
depression of about 20\%, which is modest for high current linacs.
We note that for this kind of space charge physics acceleration
can be ignored, because the only parameters that matter are the
ratio of phase advances per cell, the tune depression and the
ratios of normalized core or halo emittances. Since we always
assume that the initial emittances and phase advances in $x$ and
$y$ are equal, we only indicate ratios of $z$ and $x$ quantities
at the start.

We make the following model assumptions: (i) The initial
distribution follows the TRACEWIN option of randomly distributed
particles in a six-dimensional ellipse (here truncated at 2.8
$\sigma$); (ii)
 $10^5$ simulation particles and a 3D Poisson solver (PICNIC);
  (iii) an ``extra halo'' distribution is introduced as six-dimensional Gaussian type cut
at 2 $\sigma$ and - if present - occupied by 2.5\% of the total
number of particles; (iv) the halo size is expressed by the 99.9\%
 emittance as area of an ellipse  defined by using core twiss parameters and
  containing 99.9\% of the particles;
(v) we allow for a nonsymmetric (anisotropic) halo with
independent emittances transversely and longitudinally. The halo
emittances are described by independent multiplication factors
$M_z/M_{xy}$ with respect to the core emittances. The value of
2.5\% halo intensity is relatively arbitrary and higher than what
can usually be tolerated in high power linacs. Yet it is still low
enough to ensure that the halo space charge force is only a weak
contribution, which is the main argument for it in our study.

\begin{figure}[h]
\centering\resizebox{0.45\textwidth}{!}{\includegraphics{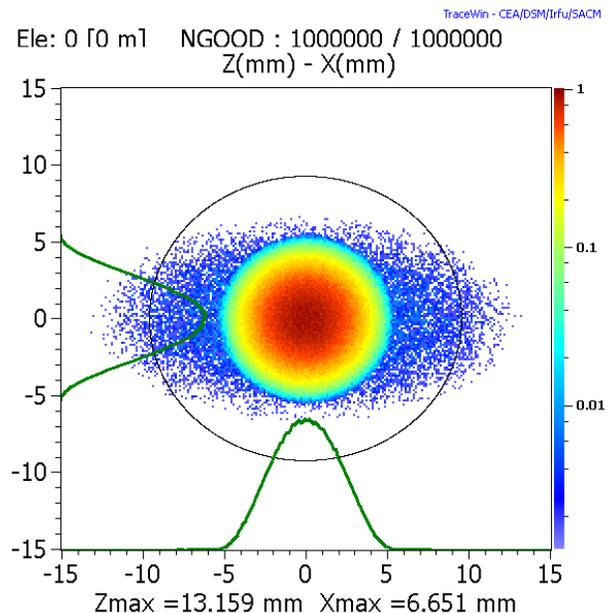}}
\caption{Exemplary initial core-halo distribution in $z-x$ plane
for halo multiplication factors $M_z/M_x=1.5/6$, with ellipse
including 99.9\% of total particles.} \label{halo}
\end{figure}

Note that this study assumes initial Gaussian halo profiles
throughout and ignores entirely the question of what physical
mechanism generates the halo, which might for example be mismatch
or errors in linear accelerators. A projection of a characteristic
initial core plus halo distribution into the real plane $z-x$ is
shown in Fig.~\ref{halo} for $M_z/M_x=6/1.5$.
%\end{document}
\section{Simulation results for dynamical crossing}\label{sec:3}
In this section we present simulation results under the assumption
of a linearly decreasing transverse phase advance such that the
 $k_z/k_x =1$ is crossed from left to right. We
mention that the analogous crossing in the opposite direction
yields similar but not equal results. There are some subtle
differences, however, which will be discussed further in
Section~\ref{sec:4}.
\subsection{Crossing without extra halo}\label{subsec:3-1}
\begin{figure}[h]
\centering\resizebox{0.45\textwidth}{!}{\includegraphics{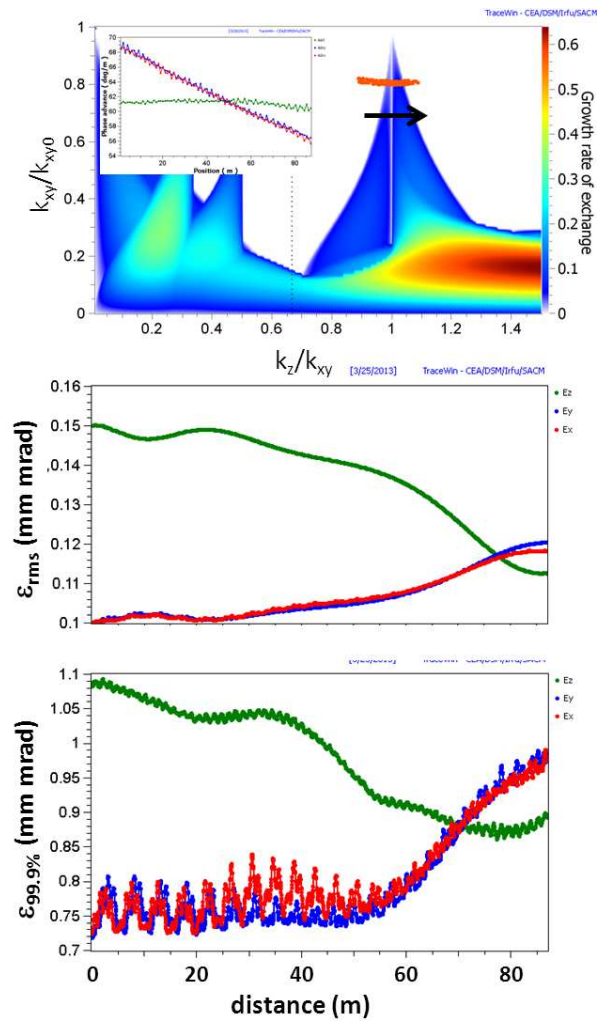}}
\caption{Crossing of main resonance from left to right in the
absence of extra halo: Tune footprint on stability chart for
$\epsilon_z/\epsilon_x =1.5$ (top,  with graphs for $k_z$ and
$k_{xy}$ in the upper left corner)), Rms (center) and 99.9\%
(bottom) emittances. } \label{firstex}
\end{figure}

For reference we first review this crossing in the absence of
extra halo as shown in Fig.~\ref{firstex} ($M_z/M_{xy}=1/1$). The
tune footprint with initial rms emittances $\epsilon_z/\epsilon_x
=1.5$  is plotted on the stability chart in the top of
Fig.~\ref{firstex}. In this chart for an initial rms emittance
ratio $\epsilon_z/\epsilon_x =1.5$ the stop-band increases with
space charge, which provides the driving term. For a detailed
discussion see Section~\ref{sec:5}). The footprint is determined
in TRACEWIN as ratio of  the $k_{z}$ and $k_{xy}$ shown in the
insert in the upper left corner of Fig.~\ref{firstex}; these
values are extracted
 as rms values of tunes obtained from tracking
of
 all particles. A linear tune sweep $k_{xy0}: 85^0 \rightarrow
70^0$ over 85 cells (here 85 m) and initial tune depression
$k_{x}/k_{0x}=0.8$ are assumed. The crossing of the center of
resonance  condition $k_{z}/k_{xy}=1$ in this case occurs at cell
50. Note that in this and all following graphs the distance on the
abscissa in meters is identical with the cell number for our
lattice example.

%%%%%%%%%%%%%% T\equiv \frac{\epsilon_z k_z}{\epsilon_x k_x},
It is noted that the crossing leads to the expected rms emittance
transfer, where the final state is nearly ``equipartitioned''. The
longitudinal rms emittance reduction is about double the increase
of each transverse one. This reflects the fact that the
longitudinal degree of freedom has to ``heat'' both transverse - a
kind of energy conservation supported by the observation that the
sum of all three rms emittances is constant within few percent.
The initial values of the 99.9\% emittances reflect a small halo
  due to truncation of the core distribution at 2.8 $\sigma$.
They follow the same pattern of exchange as the rms emittances,
including an approximate constancy of their sum. For the following
discussion it is important to note  that the equipartition process
itself does not generate additional halo - only exchange.

%%%%%%%%%%%%%%%%%%%%%%%%%%%%%%%%%%%%%%%%%%%%%%%%%%%%%%5

\subsection{Extra purely transverse halo}\label{subsec:3-2}
\begin{figure}[h]
\centering\resizebox{0.45\textwidth}{!}{\includegraphics{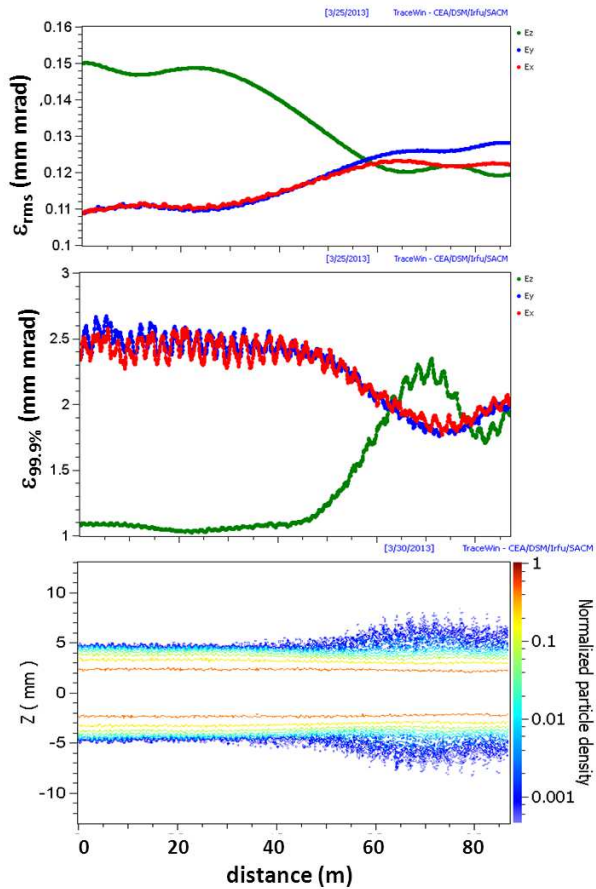}}
\caption{Extra halo added in transverse plane only: Rms (top),
99.9\% (bottom) emittances (center) and projection of
$z$-evolution (bottom).} \label{secex}
\end{figure}

Next we consider the same parameters for the core as before, but
now we add a purely transverse halo distribution with
multiplication factors $M_z/M_{xy}=1/4.5$, which results in an
initial ratio $\epsilon_z/\epsilon_x =0.4$ for the 99.9\%
emittances and inverts the core and halo emittance ratios. The
tune ramp is extended to $k_{x0}=85^0 \rightarrow 65^0$ over again
the 85 cells. The rms emittances again roughly equipartition. This
also applies to the 99.9\% emittances, but their transfer occurs
into the opposite direction: the two initially larger transverse
halo emittances ``pump'' the longitudinal halo emittance, which
more than doubles in a first swing. Thereafter, apparently, a
small fraction of halo particles oscillates back and forth between
planes - approximately around halo emittance equipartition as
shown in Fig.~\ref{secex}.

\subsection{Nearly equipartitioned core and purely longitudinal
halo} \label{subsec:3-3} 
\begin{figure}[h]
\centering\resizebox{0.45\textwidth}{!}{\includegraphics{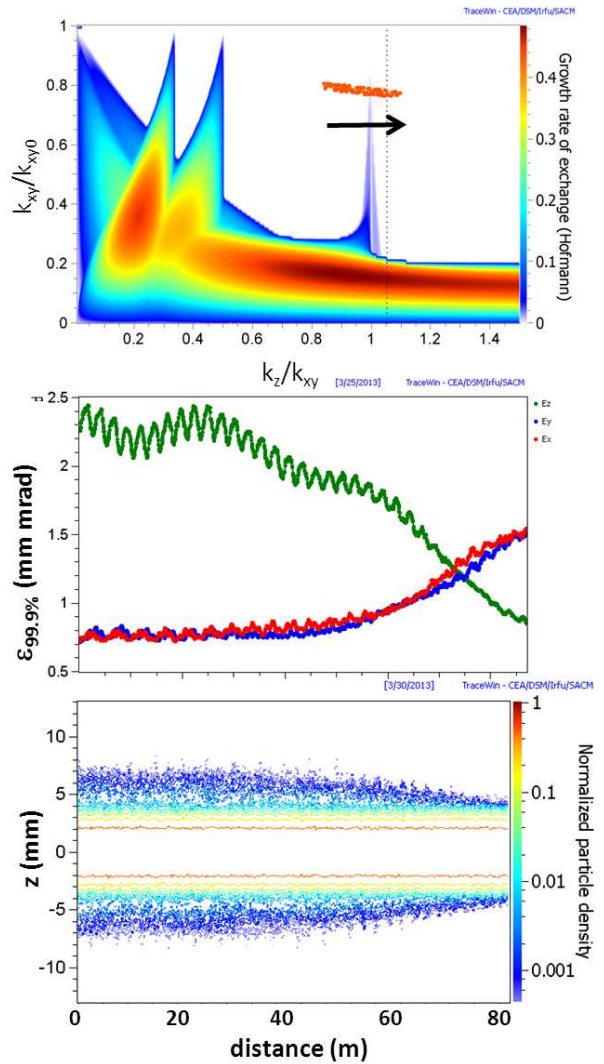}}
\caption{Extra halo in longitudinal plane with nearly
equipartitioned core: Tune footprint on stability chart for
$\epsilon_z/\epsilon_x =0.95$ (top); Rms  and 99.9\% emittances
  (middle) and  $z$-evolution (bottom).} \label{thirdex}
\end{figure}

Here we assume an almost equipartitioned
core with $\epsilon_z/\epsilon_x =0.95$, which results in a
practically negligible stopband width for the main resonance as
shown in Fig.~\ref{thirdex}. We furthermore add a purely
longitudinal halo with corresponding factors $M_z/M_{xy}=4.5/1$,
which implies a starting ratio $\epsilon_z/\epsilon_x =3$ for the
99.9\% emittances.  The transverse tune is ramped from
$k_{x0}=85^0 \rightarrow 70^0$. The coupling has a weak effect on
the rms emittances, which merge together as well - partly due to
the shrinking longitudinal halo. The 99.5\% emittances come to a
full exchange with some overshoot and qualitatively the same
behavior as the rms and 99.5\% emittances in Fig.~\ref{firstex}.
Hence, the longitudinal halo is almost fully converted into a
transverse halo with little effect on the core emittances.
Fig.~\ref{thirdex} also suggests that the coupling of halo
emittances is subject to a much broader stopband than the one
shown for the core, which is calculated for the core emittance
ratio (see also Section~\ref{sec:6}.
\subsection{Dynamical halo cleaning with transverse scrapers} \label{subsec:3-4}
Here we assume an isotropic  halo with $M_z/M_{xy}=4.5/4.5$. As
expected no emittance exchange occurs, as this
 beam  is equipartitioned both in core and halo. We next define a transverse scraper by
 introducing in three consecutive cells
 circular apertures at positions, where the beam is round. In our
 example such a scraper of 4 mm radius sharply eliminates the
 transverse halo particles (with 3\% reduction of beam intensity).
 Assuming that the scraper is located in the lattice before the coupling resonance is reached, the latter can pump the
 longitudinal halo into the transverse plane.
 The location of this scraper is found to be most effective at cells 38/39/40 (equal to distance in m), where
 $k_{z}/k_{x}=0.93$. Note that the tune ratio reaches the center of the
 stopband ($k_{z}/k_{x}=1$)
 after slightly less than 20 additional cells.
 For our
 parameters 20 cells is also the characteristic emittance exchange
 time for a static tune placed right on the stopband, hence
 without ramping across it. The issue of characteristic emittance
 exchange time is further discussed in Section~\ref{sec:4} as it
 is the basis for choosing the scraper position most effectively.

\begin{figure}[h]
\centering\resizebox{0.45\textwidth}{!}{\includegraphics{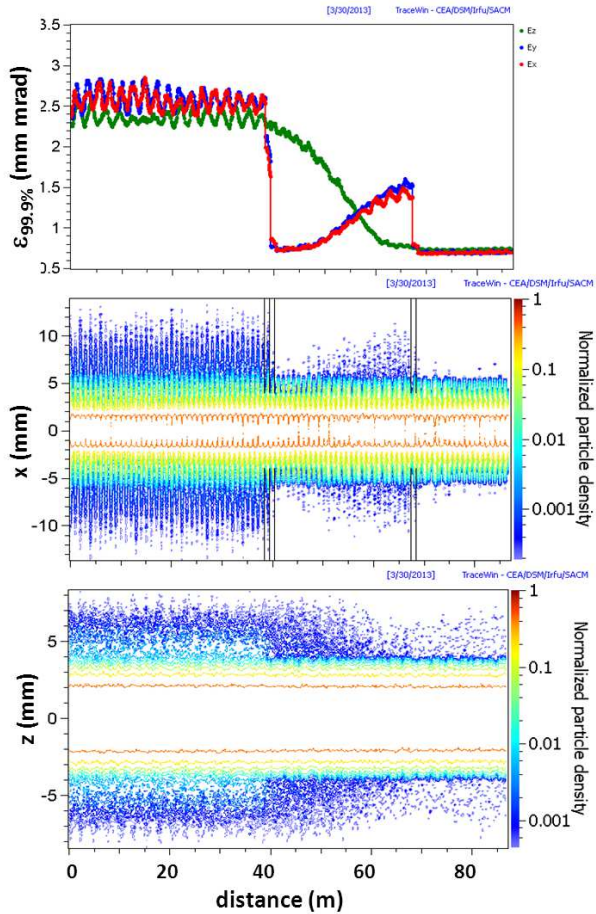}}
\caption{Cleaning of initially isotropic halo  by two consecutive
transverse scrapers: 99.9\% emittances (top); $x$-evolution
(center) and $z$-evolution (bottom).} \label{fifthex}
\end{figure}

 The result of this
 simulation is shown in
Fig.~\ref{fifthex}. It is seen that the coupling of the
longitudinal halo into the transverse plane makes the transverse
99.9\% emittances grow - following again the already observed
conservation of the sum of all three emittances. To avoid a
partial return into the longitudinal plane, which would bring the
99.9\% $\epsilon_z$ back to about the value 1.5 in our example, we
have added a second scraper with only two apertures in cells
75/76. The second scraper makes sure that the halo cleaning
remains conserved for all three planes. The rms emittances are
only changing by a few percent as a result of the scraping.
Note that the second scraper is unnecessary, if fine tuning of the
tune ramp  is realized such that the coupling stopband is left
after the longitudinal halo emittance has reached its minimum.
This helps reducing the scraped intensity - also in view of the
fact that some remaining transverse halo is not an issue of big
concern  in   linacs with large enough transverse aperture.
Removing the longitudinal halo this way appears promising as it
helps shifting the synchronous phase towards higher acceleration
rate.

\section{Simulation of static tunes}\label{sec:4}
In this section we study the topology of core and halo coupling at
the main resonance stopband for fixed values of $k_{xy0}$
(``static tunes''), which reveals some interesting details not
resolvable with dynamical crossing. In particular, the validity of
different stopband widths for core and halo is clearly seen in
cases where their emittance ratios differ.

\subsection{Same emittance ratio in core and halo}\label{subsec:4-1}
\begin{figure}[h]
\centering\resizebox{0.45\textwidth}{!}{\includegraphics{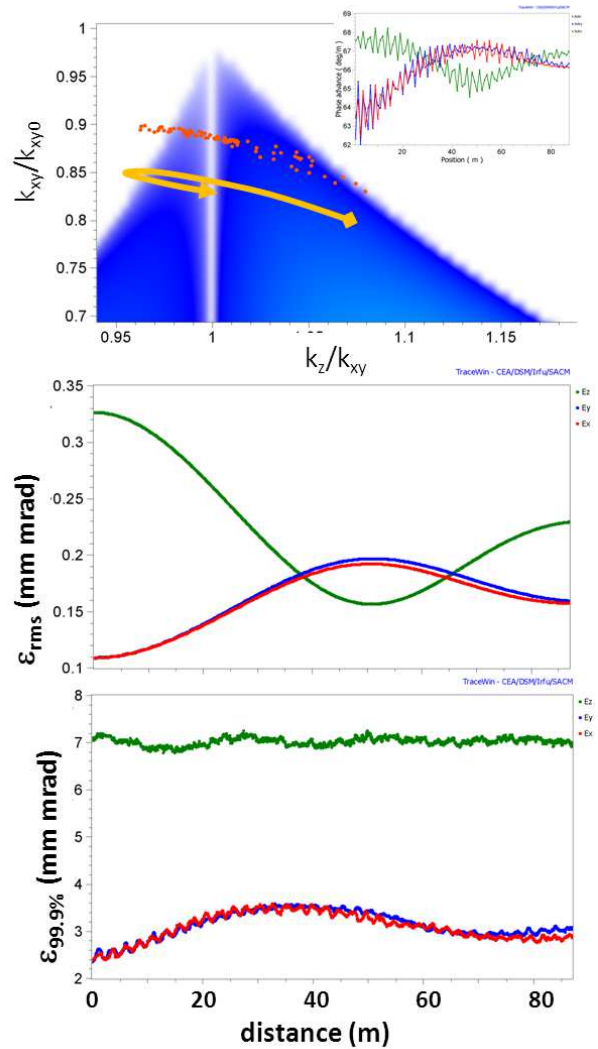}}
\caption{Selfconsistent evolution with $k_{xy0}=75^0$: tune
footprint on (zoomed) stability chart for $\epsilon_z/\epsilon_x
=3$ (top, with graphs for $k_z$ and $k_{x,y}$ in the upper right
corner); Rms (center) and 99.9\%
  (bottom) emittances.} \label{40-0}
\end{figure}

Keeping the transverse tunes fixed reveals different features of
exchange, depending on the distance from the stopband and at which
side. We assume $\epsilon_z/\epsilon_x =3$ and
$M_z/M_{xy}=4.5/4.5$, which gives the halo the same initial
emittance ratio of 3:1. With $k_{xy0}=75^0$ the corresponding
 $k_z/k_{xy}$ starts at the right edge of the stopband as shown
in Fig.~\ref{40-0}.
The starting rms emittance exchange leads to a shrinking
$\epsilon_z/\epsilon_x$, which leads to corresponding changes in
space charge densities   with the effect that $k_z/k_{xy}$ also
shrinks and moves further left into the stopband; this
self-consistent mechanism continues until equality of emittances
occurs - with even some overshoot - and $k_z/k_{xy}$ reaches the
left edge of the stopband and returns back to the center (as
indicated by an arrow in the top graph of Fig.~\ref{40-0}). The
99.9\% emittances, however, show only weak coupling.

\begin{figure}[h]
\centering\resizebox{0.45\textwidth}{!}{\includegraphics{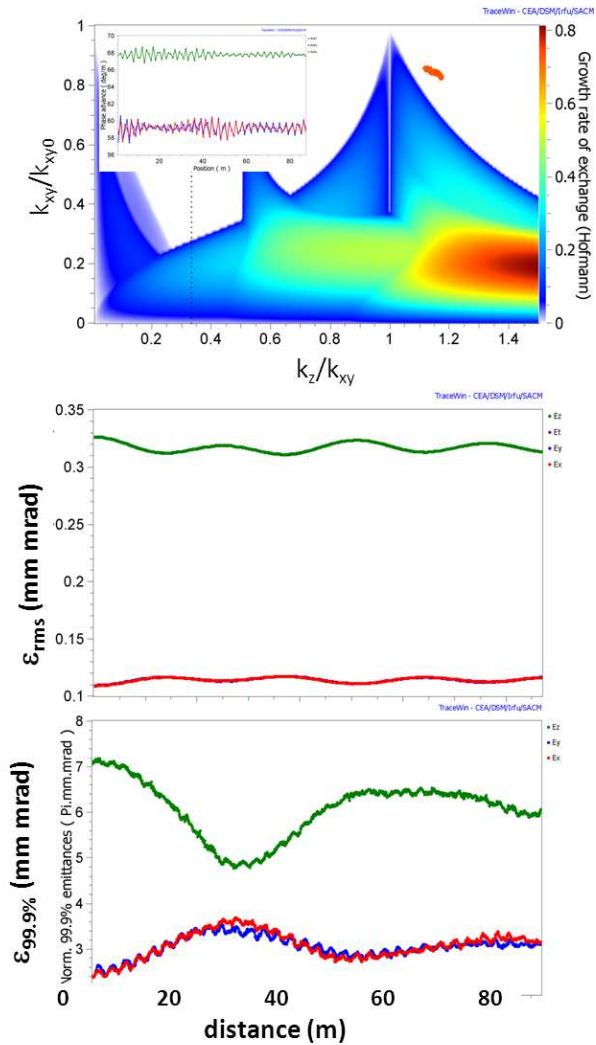}}
\caption{Selfconsistent evolution with $k_{xy0}=70^0$: tune
footprint on stability chart for $\epsilon_z/\epsilon_x =3$ (top,
with graphs for $k_z$ and $k_{x,y}$ in the upper left corner); Rms
(center) and 99.9\%
  (bottom) emittances.} \label{40-1}
\end{figure}

A very different behavior is found for $k_{xy0}=70^0$, for which
 $k_z/k_{xy}$ is actually outside the right edge of the stopband as shown in
 Fig.~\ref{40-1}.
Not surprisingly, there is practically no rms emittance exchange;
 however, the halo emittances show effective although not complete coupling.
  To interpret this phenomenon we note that tunes of halo particles differ from
  core particle tunes: the tune ratio of very large amplitude halo
  particles approaches asymptotically the space-charge-free value
  $k_{z0}/k_{xy0}$. On the other hand, the corresponding resonance condition
  $k_{z0}/k_{xy0}=1$ is satisfied close to the r.h.s. edge of the
  main resonance stopband, which helps understand the dominant halo particle coupling there.
   In fact, we have not found any effect
  on halo coupling for a working point at or beyond the l.h.s. edge
   of the main resonance stopband in Fig.~\ref{40-1}.

The asymmetric response on the main resonance stopband is also
found  in the following: for $k_z/k_{xy}$
  starting close to the left edge
of the stopband the  core emittance exchange shifts $k_z/k_{xy}$
outside of the stopband and the process stops - in contrast with
the behavior at the right edge as in Fig.~\ref{40-0}. Note that
for an initial longitudinal emittance smaller than the transverse
one the side behavior is also reversed.
\subsection{Symmetric core with longitudinal halo}\label{subsec:4-2}
\begin{figure}[h]
\centering\resizebox{0.45\textwidth}{!}{\includegraphics{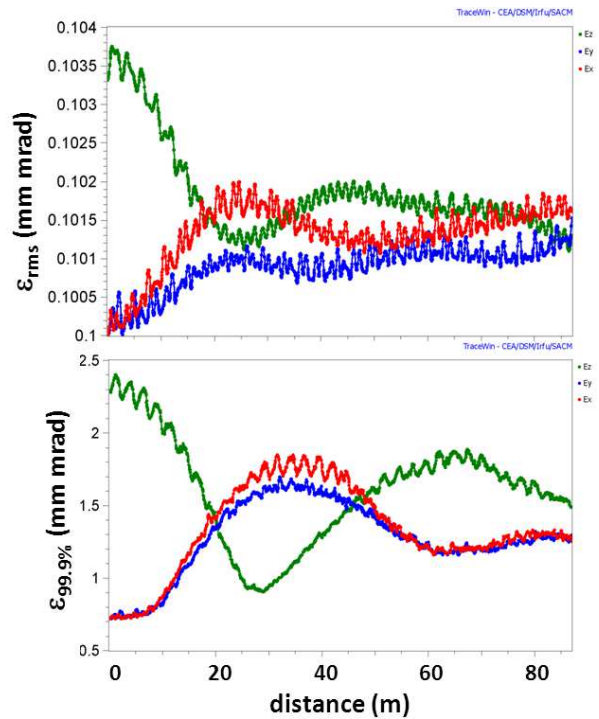}}
\caption{Static tune with $k_{z0}/k_{x0}=1$: Rms (top) and 99.9\%
  (bottom) emittances.} \label{47}
\end{figure}

Assuming $\epsilon_z/\epsilon_x =0.95$ we expect practically no
effect on the rms emittances at or near $k_z/k_{xy}=1$ as already
found in the crossing case of Fig.~\ref{thirdex}.  A purely
longitudinal halo with $M_z/M_{xy}=4.5/1$ shows emittance exchange
depending on the value of $k_z/k_{xy}$. We find that   halo
emittance coupling is most effective at the condition
$k_{z0}/k_{x0}=1$ (see Fig.~\ref{47}) and gradually vanishes away
from it in either direction. Note that this is supported by the
fact that for equal rms emittances the condition $k_z/k_{xy}=1$
coincides with the condition $k_{z0}/k_{x0}=1$ - the asymptotic
resonance condition for very large amplitude halo particles - in
contrast with the case of unequal emittances in
Section~\ref{subsec:4-1}.

By testing different values of $k_z/k_{xy}$ we actually find that
the region of halo emittance exchange is  described quite well by
a stability chart generated with the halo emittance ratio (rather
than core emittance ratio). This can be explained by the
observation that the driving term for both the exchange of core
and halo particles is the space charge multipole. This finding
will be pursued further in Sections~\ref{sec:5} and ~\ref{sec:6}.

\section{Semi-analytical scalings}\label{sec:5}
For practical applications it may be useful to consider the
approximate scaling expressions for the main resonance stopband
width and exchange time, which have been derived in
Ref.~\cite{hofmann2006}. Using $\Delta k\equiv k_{0}-k$ and the
 tune ratio $k_{z}/k_{x}$ as
variable, the stopband width    can be written in the form
\begin{equation}\label{stopband1}
\delta\left(\frac{k_{z}}{k_{x}}\right)=\frac{3}{2}
\left(\sqrt{\frac{\epsilon_z}{\epsilon_x}}-1\right) \frac{\Delta
k_z/k_{z0}}{k_x/k_{x0}},
\end{equation}
where it is noted that the longitudinal tune depression is
(approximately) related to the transverse one according to
\begin{equation}\label{stopband2}
 \Delta k_z/k_{z0}=\frac{\Delta k_x/k_{x0}}{\sqrt{\epsilon_z
 /\epsilon_x}}.
\end{equation}
As an example we mention that for $\epsilon_z/\epsilon_x=1.5$ and
$k_x/k_{x0}=0.6$ the resulting $\delta \approx 0.18$ is in good
agreement with Fig.~\ref{firstex}.

%%%%%%%%%%%%%%%%%%%%%% end Überprüfung

A corresponding expression for the e-folding time for emittance
coupling is suggested in Ref.~\cite{hofmann2006} for the center of
the stop-band. By comparison with simulation results we find that
in the region of moderate emittance imbalance $0.5
<\epsilon_z/\epsilon_x <2$ the number of cells required to reach
the first crossing of emittances is basically independent of the
emittance ratio and found as
\begin{equation}\label{exchange}
 N_{cells} \approx \frac{1}{2}\left(\Delta k_z/k_{z0}\right)^{-1}.
\end{equation}
We also find that the coupling distance estimates in
Eq.~\ref{exchange} apply quite well to the halo exchange  in spite
of the fact that the coupling of halo particles with the core -
via its space charge pseudo-octupole - is expected to decrease
with increasing distance from the core.

\section{Extended stability charts}\label{sec:6}
\begin{figure}[h]
\centering\resizebox{0.45\textwidth}{!}{\includegraphics{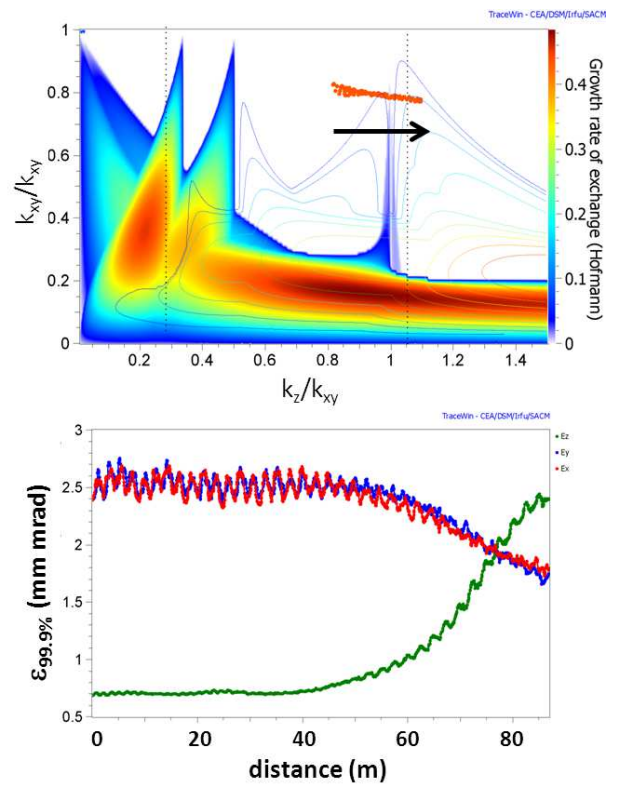}}
\caption{Crossing of main resonance: tune footprint on extended
stability chart combined for core ($\epsilon_z/\epsilon_x =0.95$)
and halo ($\epsilon_z/\epsilon_x =3.7$)(top)  and 99.9\%
    emittances (bottom).} \label{52}
\end{figure}

Our calculations give clear evidence that halo emittance exchange
may occur independent of whether a core emittance exchange takes
place or not. The conditions for effective halo emittance exchange
at the main resonance are: (i) the presence of a space charge
octupole as driving term (always given for a non-uniform beam);
(ii) fulfilment of the resonance condition and (iii) different
emittances, otherwise there is exchange of amplitudes of
individual particles without a net global transfer. It is thus
appropriate to use distinct stability charts for
 core and  halo,  where the rms emittance ratio is used for the core, and the
99.9\% (or similar) emittance ratio for the halo. This is
demonstrated in Fig.~\ref{52} for a crossing of the main resonance
from left to right with $k_{xy0}: 85^0 \rightarrow 70^0$;
furthermore the assumption of a nearly equipartitioned core
($\epsilon_z/\epsilon_x =0.95$) and a purely transverse halo with
$M_z/M_{xy}=1/4.5$, which results in a 99.9\% emittance ratio
$\epsilon_z/\epsilon_x =3.7$.  We define an extended stability
chart in the following way: the halo stability chart is plotted as
contour lines on top of the solid core chart - each one with the
corresponding emittance ratio. The 99.9\% emittance curve gives
clear evidence that this exchange occurs over a broad range of
$k_z/k_{xy}$, which we find is comparable with the width of the
stopband (halo contours) at the main resonance.

\begin{figure}[h]
\centering\resizebox{0.45\textwidth}{!}{\includegraphics{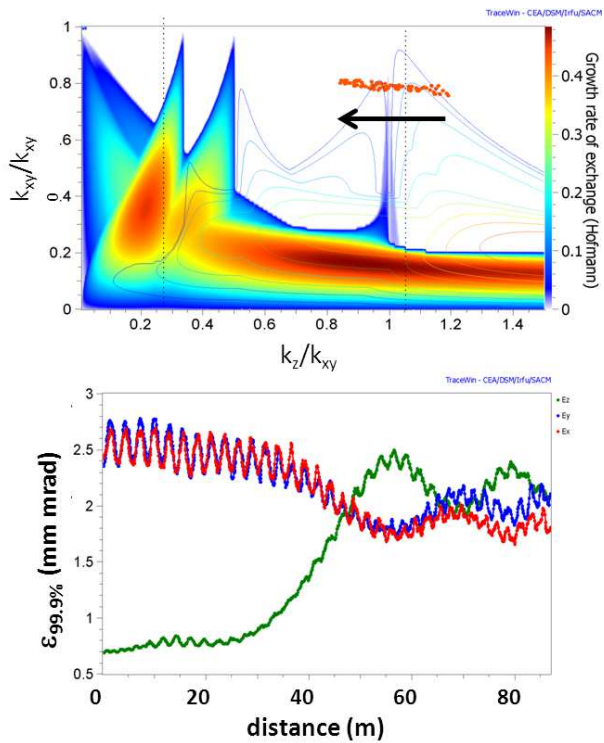}}
\caption{Same as Fig.~\ref{52} with opposite crossing of main
resonance.}  \label{53}
\end{figure}

Crossing from right to left with $k_{xy0}: 65^0 \rightarrow 85^0$
confirms a similar width of response in the variable $k_z/k_{xy}$,
which is  shown in Fig.~\ref{53}. Hence in both cases the very
narrow main resonance stopband of the core is found irrelevant for
the halo.
For creating these extended charts we have assumed that the above
described behavior at the main resonance is also applicable in an
analogous way to all other resonances on the chart.

\section{Conclusion}\label{sec:7}
We have shown that the presence of the space charge
pseudo-octupole - in all non-uniform beam models - is not only a
source of emittance exchange for the core emittances, but also for
halo emittances. The strength of this driving term is considerable
- only very few betatron periods are needed in a high current
linac to reach full exchange of emittances.

These findings allow to extend the interpretation of linac
stability charts in order to describe separately possible core and
halo emittance exchange. For the main resonance the corresponding
width is readily estimated by an approximate expression containing
the respective emittance ratio.

For practical linac design there is an incentive to avoid
undesirable transfer of halo from the transverse into the
longitudinal, which can get lost out of the bucket. This enhances
the importance of avoiding this coupling resonance - or even the
vicinity of it - in the main part of a linac. We find no
  argument at this level in support of the frequently made assumption
  that it is advantageous to make the core equipartitioned. On the contrary, it
may be appropriate to relax conditions and enable a
non-equipartitoned bunch in the interest of strictly avoiding the
resonance and halo coupling. Refraining from enforced
equipartitioning may in consequence turn out advantageous for the
design.

On the other hand, it is suggested to consider this resonance for
an active scheme of longitudinal halo transfer into the transverse
direction, where it can be scraped. In practice such use of the
resonance should be limited to a transition between linac
structures and avoided elsewhere. In any case our study makes it
clear that besides halo intensity it is also important to study
halo emittances independently for the longitudinal and transverse
planes.

%Acknowledgment: The author acknowledges ...

\end{document}